\documentclass[runningheads,a4paper]{llncs}
\usepackage{amssymb}
\usepackage{graphicx}
\usepackage{url}

\usepackage{amsfonts}
\usepackage{bm}
\usepackage{booktabs}
\usepackage{color}
\usepackage{enumitem}
  \setlist{nolistsep}
\usepackage{multirow}
\usepackage{url}

\newcommand{\R}{\mathbb{R}}

\begin{document}

\mainmatter
%================================================

%==== FILL IN ====================================
\title{Choosing a variable ordering for truth-table invariant cylindrical algebraic decomposition by incremental triangular decomposition}  % Full title
\titlerunning{Choosing a variable ordering for \texttt{RC-TTICAD}} % Shor ttitle
\author{Matthew England \and  Russell Bradford \and James H. Davenport \and David Wilson}
\authorrunning{England-Bradford-Davenport-Wilson}
\institute{
University of Bath, UK\\
\email{ \{M.England, R.J.Bradford, J.H.Davenport, D.J.Wilson\}@bath.ac.uk},\\ 
}
\maketitle

\begin{abstract}
%Must be plain text, self contained (no \cite) and < 200 words.

Cylindrical algebraic decomposition (CAD) is a key tool for solving problems in real algebraic geometry and beyond.  In recent years a new approach has been developed, where regular chains technology is used to first build a decomposition in complex space.
We consider the latest variant of this which builds the complex decomposition incrementally by polynomial and produces CADs on whose cells a sequence of formulae are truth-invariant.  
Like all CAD algorithms the user must provide a variable ordering which can have a profound impact on the tractability of a problem.  We evaluate existing heuristics to help with the choice for this algorithm, suggest  improvements and then derive a new heuristic more closely aligned with the mechanics of the new algorithm.
\end{abstract}

\section{Introduction} 
\label{SEC:Intro}

A \textit{cylindrical algebraic decomposition} (CAD) is: a \textit{decomposition} of $\R^n$, meaning a collection of cells which do not intersect and whose union is $\R^n$; \textit{cylindrical}, meaning the projections of any pair of cells with respect to a given variable ordering are either equal or disjoint; and, \textit{(semi)-algebraic}, meaning each cell can be described using a finite sequence of polynomial relations.
The original CAD by Collins \cite{ACM84I} was introduced as a tool for quantifier elimination over the reals.  Since then CAD has also been applied to problems including epidemic modelling \cite{BENW06}, parametric optimisation \cite{FPM05}, theorem proving \cite{Paulson2012}, motion planning \cite{WDEB14} and reasoning with multi-valued functions and their branch cuts \cite{DBEW12}.

\vspace*{10pt}

Traditionally, a CAD is built \textit{sign-invariant} with respect to a set of polynomials such that each one has constant sign in each cell, meaning only one sample point per cell need be tested to determine behaviour.  Collins' algorithm works in two phases.  In the \textit{projection} phase an operator is repeatedly applied to polynomials each time producing a set in one fewer variables.  Then in the \textit{lifting} phase CADs of real space are built incrementally by dimension according to the real roots of these polynomials.
%First the real line is decomposed according to the real roots of the univariate polynomials.  Then $\mathbb{R}^2$ is decomposed by repeating the process over each cell from the real line using the bivariate polynomials evaluated at a sample point, and so on.  
A full description is in \cite{ACM84I} and \cite{Collins1998} summarises improvements from the first 20 years (\cite{BDEMW13} references more recent developments).

In 2009 an approach to CAD was introduced which broke with the projection and lifting framework \cite{CMXY09}.  Instead, a \emph{complex cylindrical decomposition} (CCD) of $\mathbb{C}^n$ is built using triangular decomposition by regular chains, and then real root isolation is applied to move to a CAD of $\mathbb{R}^n$.  
We can view the CCD as an enhanced projection since gcds are calculated as well as resultants. It means the second phase is less expensive than lifting since case distinction can avoid identifying unnecessary roots.
We use \texttt{PL-CAD} for CADs built by projection and lifting and \texttt{RC-CAD} for CADs built with the new approach.
The initial work was improved in \cite{CM12b} by introducing purpose-built algorithms to refine a CCD incrementally by constraint whilst maintaining cylindricity and recycling subresultant calculations.  
A modification of the incremental algorithm to work with relations instead of polynomials then allowed for simplification in the presence of \emph{equational constraints} (ECs): equations whose satisfaction is logically implied by the input.  
The output was no longer sign-invariant for polynomials but \textit{truth-invariant} for a formula (the conjunction of relations).  Similar ideas had been developed for \texttt{PL-CAD} \cite{McCallum1999} but were difficult to generalise to multiple ECs.

In \cite{BCDEMW14}, a new variant of \texttt{RC-CAD} was presented.  Here, instead of building a CAD for a set of polynomials or relations we build one for a sequence of quantifier free formulae (QFFs) such that each formula has constant truth value on each cell: a \textit{truth-table invariant} CAD or TTICAD.  It followed the development of TTICAD theory for \texttt{PL-CAD} (see \cite{BDEMW13}, \cite{BDEMW14}) and combined it with the benefits of \texttt{RC-CAD}.  
The CCD is built using a tree structure incrementally refined by constraint.  ECs are dealt with first, with branches refined for other constraints in a formula only if the EC is satisfied.  
Further, when there are multiple ECs in a formula branches can be removed when the constraints are not all satisfied.  See \cite{BCDEMW14} and \cite{CM12b} for full details.
Building a TTICAD is often the best way to obtain a truth-invariant CAD for a single formula (if the formula has disjunctions then treating each conjunctive clause as a subformula allows simplification in the presence of any ECs) but is also the object required for applications like simplification of complex functions via branch cut analysis (see \cite{BD02} \cite{EBDW13}).
The implementation of \cite{BCDEMW14} in the \texttt{RegularChains} Library \cite{RC} (denoted \texttt{RC-TTICAD}) is our topic here.

\vspace*{10pt}

All CAD algorithms require the user to specify an ordering on the variables.  For \texttt{PL-CAD} this determines the order of projection and thus the sequence of Euclidean spaces considered en-route to $\R^n$.  For \texttt{RC-CAD} if determines both the triangular decompositions performed and the refinement to $\R^n$.
Depending on the application there may be a free or constrained choice.  For example, in quantifier elimination we must order the variables as they are quantified but may change the ordering within quantifier blocks.  
Problems easy in one variable ordering can be infeasible in another, with \cite{BD07} giving problems where one ordering leads to a cell count constant in the number of variables and another to one doubly exponential (irrespective of the algorithm used).
Hence any choice must be made intelligently.
We write $y \succ x$ if $y$ is greater than $x$ in an ordering (noting that \texttt{PL-CAD} eliminates variables from greatest to lowest in the ordering).

We start in Section \ref{SEC:Existing} by evaluating (with respect to \texttt{RC-TTICAD}) existing heuristics for choosing the variable ordering.  Then in Section \ref{SEC:New} we suggest some extensions to improve their use before developing our own heuristic more closely aligned to \texttt{RC-TTICAD}.  We give our conclusions in Section \ref{SEC:Conclusions}.

%------------------------------------------------------------
\section{Evaluating existing heuristics}
\label{SEC:Existing}

%\subsubsection{Existing Heuristics}

In what follows we assume $f$ is a polynomial, $v$ a variable and $P$ the set of polynomials defining the input to \texttt{RC-TTICAD}.  Let $\texttt{deg}(f,v)$ be the degree of $f$ in $v$, $\texttt{tdeg}(f)$ the total degree of $f$ and $\texttt{lcoeff}(f,v)$ the leading coefficient of $f$ when considered as a univariate polynomial in $v$.  For a set let $\texttt{max}$ be the maximum value, $\texttt{sum}$ the sum of values and \texttt{\#} the number of values. 
We start by considering two heuristics already in use for choosing the variable ordering in algorithms from the \texttt{RegularChains} Library \cite{RC}.
\begin{description}
\item[Triangular:]  Start with the first criteria, breaking ties with successive ones.
\begin{enumerate}
\item Let $v^{[1]} = \texttt{max}( \{ \texttt{deg}(f, v), \, | \, f \in P \} )$.  
Then set $y \succ x$ if $y^{[1]} < x^{[1]}$.
\item Let $v^{[2]} = \texttt{max}( \{ \texttt{tdeg}(\texttt{lcoeff}(f, v)), \, | \, f \in P \mbox{ (containing $v$)}\} )$.  \\
Then set $y \succ x$ if $y^{[2]} < x^{[2]}$.
\item Let $v^{[3]} = \texttt{sum}( \{ \texttt{deg}(f, v), \, | \, f \in P \} )$. 
Then set $y \succ x$ if $y^{[3]} < x^{[3]}$.
\end{enumerate}
\item[Brown:] Start with the first criteria, breaking ties with successive ones.
\begin{enumerate}
\item Set $y \succ x$ if $y^{[1]} < x^{[1]}$ (as defined in the heuristic above).
\item Let $v^{[4]} = \texttt{max}( \{ \texttt{tdeg}(t), \, | \, \mbox{$t$ is a monomial (containing $v$) from a}$ \\ $\mbox{polynomial in $P$} \} )$.  
Then set $y \succ x$ if $y^{[4]} < x^{[4]}$.
\item Let $v^{[5]} = \texttt{\#}( \{ t, \, | \, \mbox{$t$ is a monomial (containing $v$) from a polynomial}$ \\
$\mbox{in $P$} \} )$.  
Then set $y \succ x$ if $y^{[5]} < x^{[5]}$.
\end{enumerate}
\end{description}
These both involve only simple measures on the input.  The first was implemented as part of the work for \cite{CDLMXXX11} (although not formally defined there).  It is used for various triangular decomposition algorithms in the \textsc{RegularChains} Library (and is the default for \texttt{SuggestVariableOrder}). The second was developed for CAD as described in the tutorial notes \cite{Brown2004}.  In \cite{HEWDPB14} it was shown to do well in choosing a variable ordering for \textsc{Qepcad} (an implementation of \texttt{PL-CAD}). %\cite{Brown2003b}).

The next two heuristics were developed for \texttt{PL-CAD} and work by running the projection phase for each possible variable ordering and picking an optimal ordering using a measure of the projection set.
Our implementations use the projection polynomials generated by McCallum's operator \cite{McCallum1998} on $P$.
\begin{description}
\item[Sotd:] Select the variable ordering with the lowest \emph{sum of total degrees} for each of the monomials in each of the polynomials in the projection set.
\item[Ndrr:] Select the variable ordering with the lowest \emph{number of distinct real roots} of the univariate projection polynomials
\end{description}  
Sotd was suggested in \cite{DSS04} where it was found to be a good heuristic for both CAD and QE by CAD using \textsc{Redlog} (another implementation of \texttt{PL-CAD}).  % \cite{DS97a}).
Ndrr was suggested in \cite{BDEW13} as an alternative which would discriminate between differences in the real geometry. % (as opposed to the geometry in complex space).
These heuristics are clearly more expensive but note that the lifting phase does the bulk of the work for \texttt{PL-CAD}, with the projection phase often trivial (and if not then the lifting phase is likely infeasible).  %Ndrr requires real root isolation as well and so is considerably more expensive than sotd.  
%Hence when evaluating these heuristics we must weight the costs of running the heuristics against any time savings from a good choice.

%\subsubsection*{Example set}

To evaluate the heuristics we generated 600 random examples, each with two QFFs themselves a conjunction of two constraints.  
There were 100 for each of six system types: \textbf{00}, \textbf{10}, \textbf{20}, \textbf{11}, \textbf{12}, \textbf{22}.  Each digit in these labels determines to the number of ECs in a QFF (with the other constraints strict inequalities).  The polynomials defining the constraints were sparse and in three variables, generated using \textsc{Maple}'s \texttt{randpoly} function.  
\texttt{RC-TTICAD} was applied to build CADs for the problems using each of the six possible variable orderings. 
A time out of 12 minutes a problem was used affecting only six examples (one with system type \textbf{20}, two with \textbf{10} and three with \textbf{00}).  For the others, the cell count and computation time (in seconds) for each CAD was recorded.  

Table \ref{tab:All} summarises this data, showing the average and median values for each system.  First we see that (as expected) \texttt{RC-TTICAD} does better when a QFF has an EC.
%, or more accurately, does significantly worse when a QFF has no ECs. 
Although, we note the anomaly between systems of \textbf{10} and \textbf{20}.  It seems the savings from truncating branches where ECs are not simultaneously satisfied are wiped out by the costs of doing this.  It is likely the savings would be restored in the QFFs contained further non-ECs which require more processing per branch.  This will be investigated further in the future.

Next we note that the median cell counts and timings are considerably less than the mean average for every system type, indicating the presence of outliers.  We provided a third piece of data: the median of the values for each problem when averaged over the six possible orderings.  This will still avoid outlier problems but not outlier orderings.  In every case this value is much closer to the mean average, indicating that most outlying data comes from bad orderings rather than bad problems, and thus highlighting the practical importance of the ordering.

\begin{table}
\vspace*{-10pt}
\caption{Comparing \texttt{RC-TTICAD} by system type.}
\label{tab:All}
\centering
\begin{tabular}{cllllll}
\multirow{2}{*}{\textbf{System}} & \multicolumn{3}{c}{\textbf{Cell Count}} & \multicolumn{3}{c}{\textbf{Computation Time}} \\ 
\cmidrule(lr){2-4}
\cmidrule(lr){5-7}
           & Mean  & Median     & Median of av. & Mean & Median & Median of av. \\
\midrule 
\textbf{22} & 750.13  & \, 478    & \, 612.67            & 1.84    & \, 1.37   & \, 1.58 \\
\textbf{12} & 934.42  & \, 682    & \, 861.50            & 2.73    & \, 2.12   & \, 2.47 \\
\textbf{11} & 1355.45 & \, 839    & \, 1212.33           & 3.41    & \, 2.10   & \, 2.99 \\
\textbf{20} & 3271.51 & \, 2193   & \, 2918              & 8.90    & \, 6.02   & \, 7.92 \\
\textbf{10} & 2949.02 & \, 1528   & \, 2275              & 8.44    & \, 4.71   & \, 6.62 \\
\textbf{00} & 9838.76 & \, 4874   & \, 8566.67           & 34.46   & \, 17.05  & \, 29.88 
\end{tabular}
\vspace*{-15pt}
\end{table} 

%\subsubsection*{Evaluating the heuristics}

We performed the following calculations for each problem and each heuristic:
\begin{enumerate}
\item Calculate the average cell count and timing for the problem from the six possible variable orderings.
\item Run and time each heuristic for choosing a variable ordering for the problem.
\item Record the cell count and timing of the heuristic's choice.  In the event that a heuristic cannot choose between multiple orderings we take the first of their choices lexicographically (equivalent to a random choice for these problems).  
\item Calculate the saving in cell counts made by using the heuristic's choice compared to the problem average, i.e. $(1) - (3)$.  
We calculate similarly for timings but include the cost of running the heuristic, i.e. $(1) - (2) - (3)$.
\item Evaluate the savings as percentages of the problem average (so all problems are on the same scale), i.e. $100(4)/(1)$.
\end{enumerate}
Table \ref{tab:PercentageSavings} (the first four rows) shows averages of the values in (5) over problems of the same system type and the whole problem set.  All four existing heuristics offer significant cell savings and so are making good selections of variable ordering.  Although Sotd offers the highest cell savings overall, its higher costs means the Triangular heuristic is the most time efficient.
The heuristic's costs decrease as a percentage of the CAD computation time for systems with less ECs and so Sotd can achieve a much higher saving for problems of type \textbf{00} than \textbf{22}.  But there are other differences between systems not explained by running times, such as Brown generally saving more cells than Triangular but not for systems \textbf{20}.

%------------------------------------------------------------
\section{Extensions, improvements and a new heuristic}
\label{SEC:New}

\subsubsection*{Combining measures}

In \cite{BDEW13} Ndrr was developed to help with problems where Sotd could not due to differences occurring in real space only.  Hence a logical extension is to use their measures in tandem.  We have used the same evaluation for heuristics \textbf{SN} (where Ndrr is used as a tie-break for Sotd) and \textbf{NS} (where Sotd is the tie-break) with results given in Table \ref{tab:PercentageSavings}.
In both cases the tie-breaker gives marginally higher cell savings than using the single heuristic, with NS giving the highest cell saving so far, but Brown remaining the most efficient for computation time.  The costs of running these heuristics will be higher than using the single measure (at least for problems where the first measure tied) but these extra costs are usually less than the extra time savings obtained.

\vspace*{-10pt}

\subsubsection*{Greedy heuristics}

A \emph{greedy} variant of the Sotd heuristic was also suggested in \cite{DSS04} with the variable ordering decided alongside the projection phase.  At each step the projection operator is evaluated with respect to all unallocated variables and the variable whose set has lowest sum of total degree of each monomial of each new polynomial is fixed in the ordering.  We denote this heuristic \textbf{GS}.  In Table \ref{tab:PercentageSavings} we see it offers less cell savings than full Sotd (while still competitive) but has lower costs and so gives more time savings (although not as many as the Brown heuristic).  
The cost of Sotd will increase alongside the number of admissible variable orderings and so for such problems the greedy variant may offer the only sensible approach.
A greedy variant of Ndrr is not possible since that measure is on the univariate polynomials only.

\vspace*{-10pt}

\subsubsection*{Using information from \textbf{\texttt{PL-TTICAD}}}

The projection sets used so far are those for a sign-invariant \texttt{PL-CAD}, thus considering not the constraints in the input but the polynomials defining them.  Since \texttt{RC-TTICAD} is building a TTICAD (usually smaller for all except systems \textbf{00}), a sensible extension is to use the projection phase from \texttt{PL-TTICAD} \cite{BDEMW14}.  However, we cannot match the declared output structure exactly: \texttt{PL-TTICAD} uses (at most) one declared EC per QFF (with others treated the same as non-ECs).  Hence, for QFFs with 2 ECs we will run the projection phase with the first of these declared (so for example, systems \textbf{20} are treated the same as \textbf{10}).  We denote the heuristics applying Sotd and Ndrr with this projection set as \textbf{S-TTI} and \textbf{N-TTI}.  From Table \ref{tab:PercentageSavings} we see they offer substantially more cell savings than their standard versions.  They also achieves higher time savings: both due to the improved choices and lower running costs (since the TTICAD projection operator is a subset of the sign-invariant one).

We can also run the greedy variant of Sotd with the \texttt{PL-TTICAD} projection phase (denoted \textbf{GS-TTI}.  In Table \ref{tab:PercentageSavings}) we see this decreases the cell savings on offer by S-TTI (slightly) but that the decreased cost increases the time savings.  

\subsubsection{Developing a new heuristic}

We now aim to develop a new heuristic, which considers more algebraic information than the input but is tailored to \texttt{RC-TTICAD} itself rather than a \texttt{PL-CAD} algorithm.  
The main saving offered by the regular chains approach is case distinction meaning that not all projection factors are considered universally.  For example, the second coefficient in a polynomial is only considered when the first vanishes (and then only evaluated modulo that constraint).  Consider a set of polynomials consisting of the following:
\begin{itemize}
\item the discriminants, leading coefficients and cross-resultants of the polynomials forming the first constraint in each QFF;
\item if a QFF has no EC then also the (other) discriminants, leading coefficients and cross resultants of all polynomials defining constraints there;
\item if a QFF has more than one EC then also the resultant of the polynomial defining the first with that of the second.
\end{itemize}
Here the resultants, discriminants and coefficients are taken with respect to the first variable in the ordering.  We can observe that these polynomials will all be sign-invariant in the output.  See \cite{BCDEMW14}, \cite{CM12b} for the algorithm specifications and \cite{EBCDMW14} for a fuller discussion and examples of this (from a study in the context of choosing the constraint ordering).  This set of polynomials does not contain all those computed by \texttt{RC-TTICAD}.
Their importance is that they are considered in their own right rather than modulo others.  

We define a new heuristic to pick variable orderings in two stages:  First the polynomials forming the input are considered and variables ordered according to maximum degree (as with Triangular and Brown).  Then ties are broke by calculating the set of polynomials described above for each unallocated variable and setting $y \succ x$ if the maximum degree in $y$ is less than $x$.  We denote this \textbf{NewH} and in Table \ref{tab:PercentageSavings} we see it achieves almost as many savings as S-TTI despite using a smaller set of polynomials to make its choice.  

We could go further by including some more of the missing information.  For example, we can use the degree of the omitted discriminants, resultants and leading coefficients as a third tie-break.  This heuristic is denoted \textbf{NewH-ext} and the results of its evaluation are in the final row of Table \ref{tab:PercentageSavings}.  We see it achieves even higher cell savings (and the greatest time savings of any heuristic).

%------------------------------------------------------------
\section{Conclusions}
\label{SEC:Conclusions}

We have demonstrated that the variable ordering can have a great effect on \texttt{RC-TTICAD} and using any of the heuristics discussed is much better than a random choice.  Simple measures on the input can be effective, but more cell savings can be obtained by using information from the projection phase of \texttt{PL-CAD}.  This can be extended to time savings if we use a greedy variant or a projection phase more aligned to the system type.  
We have also suggested a new heuristic aligned to \texttt{RC-TTICAD} which achieves similar savings with less information by identifying a set of polynomials of importance to the algorithm.  It was sufficient for allocating two variables (and hence ordering three) as required by our problem set.  Extending to problems with more variables is a topic of future work.  

The heuristics performance varied with the system classes and so heuristics that changed along with this performed better.  The precise relationships at work here are not always clear to see.  Machine learning on the set of measures used by the heuristics may offer a meta-heuristic greater than the sum of its parts (as was found to be the case recently when choosing a variable ordering for \textsc{Qepcad} \cite{HEWDPB14}).  
Finally, we note that when using \texttt{RC-TTICAD} there are questions of problem formulation other that the variable ordering to use.  As implied in Section \ref{SEC:New}, the order the constraints are presented affects the output.  Advice on making this choice intelligently was derived in \cite{EBCDMW14}.

\begin{table}
\vspace*{-15pt}
\caption{Comparing the savings (as a percentage of the problem average) in cells (C) and net timings (NT) from various heuristics.}
\centering
\label{tab:PercentageSavings}
\begin{tabular}{lllllllllllllll}
\multirow{2}{*}{\textbf{Heuristic}}
& \multicolumn{2}{c}{\textbf{22}} 
& \multicolumn{2}{c}{\textbf{12}} 
& \multicolumn{2}{c}{\textbf{11}} 
& \multicolumn{2}{c}{\textbf{20}} 
& \multicolumn{2}{c}{\textbf{10}} 
& \multicolumn{2}{c}{\textbf{00}} 
& \multicolumn{2}{c}{\textbf{All}}  \\
\cmidrule(lr){2-3}
\cmidrule(lr){4-5}
\cmidrule(lr){6-7}
\cmidrule(lr){8-9}
\cmidrule(lr){10-11}
\cmidrule(lr){12-13}
\cmidrule(lr){14-15}
           & \,\, C & \, NT & \,\, C & \, NT & \,\, C & \, NT 
           & \,\, C & \, NT & \,\, C & \, NT & \,\, C & \, NT & \,\, C & \, NT \\
\midrule 
\textbf{Triangular}  & 32.6 & 33.9 & 33.9 & 34.0 & 40.9 & 41.3 & 47.9 & 46.8 & 47.7 & 47.2 & 56.0 & 58.8
   & 43.0 & 43.6 \\
\textbf{Brown}  & 37.6 & 39.1 & 39.3 & 39.8 & 45.9 & 47.1 & 45.0 & 44.3 & 51.6 & 50.9 & 61.9 & 64.5
   & 46.8 & 47.5 \\
\textbf{Sotd}  & 36.7 & 23.9 & 37.9 & 27.7 & 49.4 & 40.4 & 42.8 & 39.5 & 56.3 & 53.9 & 59.9 & 61.8
   & 47.1 & 41.0 \\
\textbf{Ndrr}  & 40.1 & 21.2 & 44.1 & 33.0 & 40.2 & 30.7 & 35.7 & 34.4 & 54.8 & 51.3 & 54.0 & 54.3 
   & 44.9 & 37.4 \\
\midrule
\textbf{SN} &37.0 & 24.3 & 37.2 & 27.4 & 49.2 & 40.4 & 42.5 & 39.6 & 56.0 & 53.5 & 60.4 & 62.5
   & 47.0 & 41.1   \\
\textbf{NS} & 41.3 & 22.6 & 41.2 & 30.7 & 47.8 & 37.1 & 38.7 & 36.0 & 57.1 & 51.7 & 58.4 & 60.2
   & 47.3 & 39.6   \\  
\midrule 
\textbf{GS} & 35.0 & 32.7 & 33.7 & 32.5 & 49.5 & 46.5 & 39.8 & 38.9 & 52.3 & 52.1 & 52.5 & 55.9 
   & 43.8 & 43.3   \\
\midrule
\textbf{S-TTI} & 42.7 & 40.4 & 46.4 & 43.2 & 55.0 & 49.1 & 48.4 & 48.1 & 61.2 & 60.2 & 59.9 & 61.7
   & 52.2 & 50.3 \\ 
\textbf{N-TTI} & 48.5 & 37.1 & 46.8 & 40.5 & 48.6 & 42.3 & 47.8 & 46.9 & 59.0 & 55.3 & 54.0 & 54.3
   & 50.7 & 46.0 \\   
\textbf{GS-TTI} & 46.4 & 47.2 & 44.9 & 44.5 & 56.7 & 54.7 & 49.3 & 50.2 & 56.7 & 57.5 & 52.8 & 55.9
   & 51.1 & 51.6 \\
\midrule
\textbf{NewH}   & 45.9 & 45.5 & 41.8 & 43.5 & 51.4 & 50.8 & 48.2 & 47.6 & 56.4 & 52.4 & 67.0 & 68.5
   & 51.7 & 51.3 \\
\textbf{NewH-ext} & 46.2 & 45.9 & 42.2 & 43.3 & 51.6 & 51.4 & 49.3 & 49.5 & 55.9 & 52.0 & 67.0 & 68.5
   & 52.0 & 51.7      
\end{tabular}
\vspace*{-30pt}
\end{table}

%------------------------------------------------------------

\subsection*{Acknowledgements}
This work was supported by EPSRC grant EP/J003247/1.  \texttt{RC-TTICAD} was developed by Changbo Chen, Marc Moreno Maza and the present authors.

\vspace*{-10pt}

%\bibliographystyle{plain}
%\bibliography{CAD}

\begin{thebibliography}{99}

\bibitem{ACM84I}
D.~Arnon, G.E. Collins, and S.~McCallum.
\newblock Cylindrical algebraic decomposition {I}: The basic algorithm.
\newblock {\em SIAM J. Comput.}, 13:865--877, 1984.

\bibitem{BCDEMW14}
R.~Bradford, C.~Chen, J.H. Davenport, M.~England, M.~Moreno Maza, and
  D.~Wilson.
\newblock Truth table invariant cylindrical algebraic decomposition by regular
  chains.
\newblock {\em Submitted for publication.} Preprint: \url{http://opus.bath.ac.uk/38344/}, 2014.

\bibitem{BD02}
R.~Bradford and J.H. Davenport.
\newblock Towards better simplification of elementary functions.
\newblock {\em Proc. ISSAC '02}, pages 16--22. ACM, 2002.

\bibitem{BDEMW13}
R.~Bradford, J.H. Davenport, M.~England, S.~McCallum, and D.~Wilson.
\newblock Cylindrical algebraic decompositions for boolean combinations.
\newblock {\em Proc. ISSAC '13}, pages 125--132. ACM, 2013.

\bibitem{BDEMW14}
R.~Bradford, J.H. Davenport, M.~England, S.~McCallum, and D.~Wilson.
\newblock Truth table invariant cylindrical algebraic decomposition.
\newblock {\em Submitted for publication.} Preprint: \url{http://opus.bath.ac.uk/38146/}, 2014.

\bibitem{BDEW13}
R.~Bradford, J.H. Davenport, M.~England, and D.~Wilson.
\newblock Optimising problem formulations for cylindrical algebraic
  decomposition.
\newblock In: {\em Intelligent Computer Mathematics} (LNCS 7961), pages 19--34. Springer Berlin Heidelberg, 2013.

\bibitem{Brown2004}
C.W. Brown.
\newblock Companion to the tutorial: {C}ylindrical algebraic decomposition,
  presented at {ISSAC} 2004.  Available from:
\newblock
  \url{http://www.usna.edu/Users/cs/wcbrown/research/ISSAC04/handout.pdf}.

\bibitem{BD07}
C.W. Brown and J.H. Davenport.
\newblock The complexity of quantifier elimination and cylindrical algebraic
  decomposition.
\newblock {\em Proc. ISSAC '07}, pages 54--60. ACM, 2007.

\bibitem{BENW06}
C.W. Brown, M.~El Kahoui, D.~Novotni, and A.~Weber.
\newblock Algorithmic methods for investigating equilibria in epidemic
  modelling.
\newblock {\em J. Symbolic Computation}, 41:1157--1173, 2006.

\bibitem{CDLMXXX11}
C.~Chen, J.H. Davenport, F.~Lemaire, M.~Moreno Maza, B.~Xia, R.~Xiao, and
  Y.~Xie.
\newblock Computing the real solutions of polynomial systems with the
  \textsc{RegularChains} library in \textsc{Maple}.
\newblock {\em ACM C.C.A.}, 45(3/4):166--168, 2011.

\bibitem{CM12b}
C.~Chen and M.~Moreno Maza.
\newblock An incremental algorithm for computing cylindrical algebraic
  decompositions.
\newblock {\em Proc. ASCM '12}. To appear in LNAI, Springer. Preprint: \url{http://arxiv.org/abs/1210.5543}, 2012.

\bibitem{CMXY09}
C.~Chen, M.~Moreno Maza, B.~Xia, and L.~Yang.
\newblock Computing cylindrical algebraic decomposition via triangular
  decomposition.
\newblock {\em Proc. ISSAC '09}, pages 95--102. ACM, 2009.

\bibitem{Collins1998}
G.E. Collins.
\newblock Quantifier elimination by cylindrical algebraic decomposition -- 20
  years of progress.
\newblock In: {\em Quantifier Elimination and Cylindrical Algebraic Decomposition}, Texts \& Monographs in Symb.
  Com., pages 8--23. Springer-Verlag, 1998.

\bibitem{DBEW12}
J.H. Davenport, R.~Bradford, M.~England, and D.~Wilson.
\newblock Program verification in the presence of complex numbers, functions
  with branch cuts etc.
\newblock {\em Proc. 14th SYNASC '12}, pages 83--88. IEEE, 2012.

\bibitem{DSS04}
A.~Dolzmann, A.~Seidl, and T.~Sturm.
\newblock Efficient projection orders for {CAD}.
\newblock {\em Proc. ISSAC '04}, pages 111--118. ACM, 2004.

\bibitem{EBCDMW14}
M.~England, R.~Bradford, C.~Chen, J.H. Davenport, M.~Moreno Maza, and
  D.~Wilson.
\newblock Problem formulation for truth-table invariant cylindrical algebraic
  decomposition by incremental triangular decomposition.
\newblock {\em To Appear} Proc. CICM '14 (LNAI 8543), pages 46--60, 2014.  
Preprint: \url{http://opus.bath.ac.uk/39231/}

\bibitem{EBDW13}
M.~England, R.~Bradford, J.H. Davenport, and D.~Wilson.
\newblock Understanding branch cuts of expressions.
\newblock In: {\em Intelligent Computer Mathematics} (LNCS vol. 7961), pages 136--151. Springer Berlin Heidelberg, 2013.

\bibitem{FPM05}
I.A. Fotiou, P.A. Parrilo, and M.~Morari.
\newblock Nonlinear parametric optimization using cylindrical algebraic
  decomposition.
\newblock {\em Proc. CDC-ECC '05}, pages 3735--3740, 2005.

\bibitem{HEWDPB14}
Z.~Huang, M.~England, D.~Wilson, J.H. Davenport, L.~Paulson, and J.~Bridge.
\newblock Applying machine learning to the problem of choosing a heuristic to
  select the variable ordering for cylindrical algebraic decomposition.
\newblock {\em To Appear:} Proc. CICM '14 (LNAI 8543), pages 92--107, 2014.  
Preprint: \url{http://opus.bath.ac.uk/39232/}

\bibitem{McCallum1998}
S.~McCallum.
\newblock An improved projection operation for cylindrical algebraic
  decomposition.
\newblock In: {\em Quantifier Elimination and Cylindrical Algebraic Decomposition}, Texts \& Monographs in Symb. Comp., pages 242--268. Springer-Verlag, 1998.

\bibitem{McCallum1999}
S.~McCallum.
\newblock On projection in {CAD}-based quantifier elimination with equational
  constraint.
\newblock {\em Proc. ISSAC '99}, pages 145--149. ACM, 1999.

\bibitem{Paulson2012}
L.C. Paulson.
\newblock Metitarski: Past and future.
\newblock In: {\em Interactive Theorem Proving} (LNCS vol. 7406), pages 1--10. Springer, 2012.

\bibitem{WDEB14}
D.~Wilson, J.H. Davenport, M.~England, and R.~Bradford.
\newblock A ``piano movers'' problem reformulated.
\newblock {\em Proc. SYNASC '13}, pp. 53--60. IEEE, 2014.

\bibitem{RC}
The \textsc{RegularChains} Library: \url{http://www.regularchains.org}

\end{thebibliography}

\end{document}